\documentclass[manuscript,authorversion,nonacm]{acmart}

\AtBeginDocument{%
  }

\newcommand{\accessed}{\textit{, last accessed \today}}

\usepackage{caption}
\usepackage{subcaption}

\usepackage{tabularx}
\newcommand\setrow[1]{\gdef\rowmac{#1}#1\ignorespaces}
\newcommand\clearrow{\global\let\rowmac\relax}
\clearrow 

\usepackage{acro}
\DeclareAcronym{UI}{short = UI, long = user interface}
\DeclareAcronym{GUI}{short = GUI, long = graphical user interface}
\DeclareAcronym{TLX}{short = TLX, long = NASA-Task Load Index}
\DeclareAcronym{RTLX}{short = raw TLX, long = raw NASA-Task Load Index}
\DeclareAcronym{ER}{short = ER, long = error rate}
\DeclareAcronym{TCT}{short = TCT, long = task completion time}
\DeclareAcronym{HCI}{short = HCI, long = human-computer interaction}
\DeclareAcronym{UX}{short = UX, long = user experience}
\DeclareAcronym{HFE}{short = HFE, long = Human Factors and Ergonomics}
\DeclareAcronym{AI}{short = AI, long = Artificial Intelligence}
\DeclareAcronym{cuDNN}{short = cuDNN, long =  CUDA Deep Neural Network library}
\DeclareAcronym{RMSE}{short = RMSE, long = root mean squared error}
\DeclareAcronym{RF}{short = RF, long = Random Forest}
\DeclareAcronym{GP}{short = GP, long = Gaussian process, long-plural = Gaussian processes}
\DeclareAcronym{KNN}{short = \textit{k}NN, long = \textit{k}-nearest neighbor}
\DeclareAcronym{NN}{short = NN, long = Neural Network}
\DeclareAcronym{DNN}{short = DNN, long =  Deep Neural Network}
\DeclareAcronym{CNN}{short = CNN, long = Convolutional Neural Network}
\DeclareAcronym{FCL}{short = FCL, long = fully connected layer}
\DeclareAcronym{LOOCV}{short = LOOCV, long = leave-one-out cross-validation}
\DeclareAcronym{CV}{short = CV, long = cross-validation}
\DeclareAcronym{RM}{short = RM, long = repeated measure}
\DeclareAcronym{ANOVA}{short = ANOVA, long = analysis of variance}
\DeclareAcronym{RMANOVA}{short = RM-ANOVA, long = repeated measures analysis of variance}

\usepackage[commandnameprefix=always, final]{changes}

\begin{document}


\title[Exploring the Lands Between]{Exploring the Lands Between: A Method for Finding Differences between AI-Decisions and Human Ratings through Generated Samples}


\author{Lukas Mecke}
\affiliation{%
 \institution{University of the Bundeswehr Munich}
  \city{Munich}
  \country{Germany}}
\affiliation{%
 \institution{LMU Munich}
  \city{Munich}
  \country{Germany}}
\email{lukas.mecke@unibw.de}

\author{Daniel Buschek}
\affiliation{%
 \institution{University of Bayreuth}
  \city{Bayreuth}
  \country{Germany}}
\email{daniel.buschek@uni-bayreuth.de}

\author{Uwe Gruenefeld}
\affiliation{%
 \institution{University of Duisburg-Essen}
  \city{Essen}
  \country{Germany}}
\email{uwe.gruenefeld@uni-due.de}

\author{Florian Alt}
\affiliation{%
 \institution{University of the Bundeswehr Munich}
  \city{Munich}
  \country{Germany}}
\email{florian.alt@unibw.de}

\renewcommand{\shortauthors}{Mecke et al.}

\begin{abstract}

Many important decisions in our everyday lives, \chadded{such as authentication via biometric models,} are made by \ac{AI} systems. These can be in poor alignment with human expectations 
, and testing them on clear-cut existing data may not be enough to uncover those cases. 
We propose a method to find samples in the latent space of a generative model that is challenging for a decision-making model with regard to matching human expectations. By presenting those samples to both the model and human raters, we can identify areas where its decisions align with human intuition and where they contradict it. 
We apply this method to a face recognition model and collect a dataset of 11,200 human ratings from 100 participants. We discuss findings from our dataset and how our approach can be used to explore the performance of AI models in different contexts and for different user groups.
\end{abstract}



\begin{teaserfigure}
    \centering
    \includegraphics[width=\linewidth]{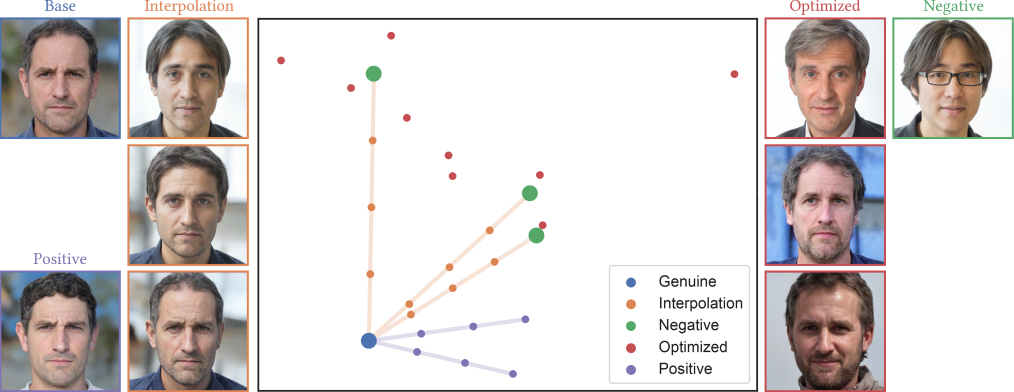}
    \caption{We propose to strategically sample alterations (positive, negative, interpolation, and optimized) of a base image from the latent space of a generative model (middle: projection of those samples on two dimensions of this latent space) to gain insights into another model that makes decisions on this data (e.g., face recognition): This involves comparing its ratings of those samples (e.g., recognition scores) to ratings by human raters. We explore this approach for the use case of examining face recognition models by sampling images from the latent space of StyleGAN2~\cite{Karras2020cvpr}.}
    \label{fig:samples}
\end{teaserfigure}

\maketitle
\newpage

\section{Introduction}

Today, more and more people are confronted with \ac{AI} decision-making systems in their everyday lives.
Applications of these systems cover a wide range, from professional (for example, supporting doctors in their diagnosis~\cite{topol2019high}) to personal scenarios \chadded{such as}\chdeleted{(for example,} identifying friends in photos taken~\cite{schroff2015facenet} \chadded{or authentication through biometric systems}\chdeleted{)}. 
However, these systems \chadded{can} make decisions that are sometimes hard to understand, surprising, inconsistent, or in little alignment with human perception or intuitions. \chadded{As an example, Pixel phones authenticated users with closed eyes, allowing for unwanted unlocks in their sleep}\footnote{\url{https://www.engadget.com/2019-10-21-google-pixel-4-face-unlock-eyes-open.html}\accessed}.
\chdeleted{As an everyday example, many modern smartphones offer the functionality to unlock them with a face scan, sometimes showing less common sense than expected, as the ability to unlock a Pixel phone with closed eyes demonstrates}
\chdeleted{Current challenges of decision-making systems are not limited to understanding them.}
\chdeleted{Across many users and decisions, }Some sensor-based decision systems may enact hidden biases learned from the training data~\cite{cabrera2019fairvis,terhorst2021comprehensive}. For example, some facial recognition systems have been reported to work worse for people of colour~\cite{cabrera2019fairvis,yapo2018ethical}. 
Furthermore, system performance can vary greatly depending on the user~\cite{yager2008biometric} or specific user groups~\cite{cabrera2019fairvis}. 
Such unfair, unexpected, and inexplicable decisions are particularly critical in \chadded{biometric systems as they often}\chdeleted{those systems that} serve as gatekeepers for access to facilities, financial and other resources, data, or devices. 

Thus, \chadded{global}\chdeleted{overall} performance metrics, like precision or recall of a system, \chadded{that are the current state of the art for describing model behavior} may not be relevant and applicable to the individual. 
Such methods of testing decision-making models can fall short, as they \chadded{do not account for decision alignment with users' expectations and} rely on real-world data with clear-cut expected decisions. \chdeleted{(i.e. it is clear if two images were taken of the same person or belong to different identities). }The model's performance and potential weaknesses may become more apparent when exploring the space between, where \chadded{it is less clear if two images belong to different identities, making decisions harder}\chdeleted{decisions are more difficult}.

Following this idea, this paper presents a method to explore the decision boundaries of (black-box) decision-making systems using artificially generated inputs.
We propose three main areas of inputs: samples that are expected to lead to a positive decision, samples that are expected to lead to a negative decision, and samples that explore the space in between where the decision may be unclear. \chadded{For authentication models like biometrics, ratings of those samples can give direct insights into their performance under different conditions.} Generating artificial samples has two main advantages: no real-world samples are needed (beyond training the generative model), and we can explore cases that would not be possible in real life to better understand \chadded{model performance under certain conditions}\chdeleted{the model} (e.g., a voice that is \chadded{modified to sound more like a different speaker}\chdeleted{a mix of two speakers}). 
By presenting these samples both to human raters and decision models, it is then possible to find areas of agreement and disagreement \chadded{that can inform the design and improvement of decision-making models and user interfaces for them}. 

As a concrete application of a \chadded{model making security-relevant decisions for its users}\chdeleted{decision-making model}, we evaluate our method by exploring face recognition with color images as input (e.g., face unlock with a phone camera). This is both easy to interpret (humans themselves are excellent at recognizing faces~\cite{diamond1986faces}) and of high societal relevance and impact\footnote{ACM statement on the topic: \url{https://www.acm.org/media-center/2022/february/tpc-tech-brief-facial-recognition}\accessed}. We generate meaningful alterations (see Figure \ref{fig:samples}) for 40 base images using StyleGAN2~\cite{Karras2020cvpr} and collected a dataset of perceived similarity and identity of the presented image pairs in an online comparison task with 100 participants \chadded{to demonstrate the viability and potential use cases for our method}. 

We find interesting mismatches between the analyzed face recognition model and human raters. The model rated images of children more similar and semantic changes (like the addition of glasses) less similar than humans would. Our optimized samples successfully fooled the decision model while mostly being perceived as different by humans. Latent distance and perceptual distance were good predictors for perceived identity except for those optimized samples. 

\chadded{Those results show that our method can provide insights into the performance of biometric models, support developers in improving them, and help to communicate a more nuanced view of model performance to end-users. We conclude with a discussion of those opportunities and the applicability of our method to other models, use cases, and user groups.}\chdeleted{
We conclude by discussing how our approach could be used to explore the performance of AI models in different contexts and for different user groups.}

\textbf{Contribution Statement.}
This work makes multiple contributions: we 1) propose a method using generated samples to \chadded{gain insights into the alignment with human intuition of}\chdeleted{understand} decision-making systems by presenting them with selected samples and comparing the provided decisions to human ratings; we \chdeleted{2) }implement this method for the example case of face recognition \chadded{and 2)}\chdeleted{, 3)} provide the generated dataset \chadded{as well as 3) }\chdeleted{, and 4) contribute} an initial analysis of this data and a discussion of further application areas of our method.

\section{Related Work}\label{sec:RW}

In this section, we focus on approaches for generating artificial content \chadded{and}\chdeleted{,} previous uses of such approaches\chadded{. We introduce methods of introspection into model performance} as well as the \chadded{related} topic of adversarial samples \chdeleted{that follow a similar idea to our proposed method}.

\subsection{\chadded{Artificial Content Generation}}
In the last years, approaches for artificial sample generation have gained public awareness through methods like ChatGPT\footnote{\chadded{\url{https://chat.openai.com}}} for text production or Stable Diffusion~\cite{rombach2022high} and DALL-E~\cite{ramesh2021zero} for generating images from text prompts. \chadded{Other approaches include}\chdeleted{Beyond those, there exists a plethora of other approaches for generating content, including} autoencoders~\cite{kingma2013auto,vahdat2020nvae,pidhorskyi2020adversarial}, normalizing flows~\cite{rezende2015variational,kingma2018glow}, and Generative Adversarial Networks~\cite{goodfellow2014generative,Karras2020cvpr,engel2018gansynth,kang2023gigagan}(GANs), to name just a few. 
\chadded{Those models learn}\chdeleted{The main principle behind all those models is to learn} a representation of the distribution of their training data (latent space) that can then be used to generate new and altered samples. In particular for GANs, there exist many approaches showing how their latent space can be used to generate semantic edits (for example, making a person older)~\cite{shen2020interpreting,jahanian2019steerability,wang2021attribute}, find meaningful dimensions~\cite{harkonen2020ganspace}, and mixing samples both on a style and content level~\cite{Karras2020cvpr,choi2018stargan}. Beyond artistic purposes, they can also be used to generate synthetic data for training and evaluating machine learning models~\cite{karras2019style,colbois2021use} and finding biases~\cite{denton2019image,balakrishnan2021towards}. With our approach, we utilize those models to produce challenging samples for a decision-making model. \chadded{Given the recent advances in the field, we believe that our method is (or soon will be) applicable to many other (security) fields (e.g. speaker recognition using synthetic voice ~\cite{ning2019review}) as well.} 

A related approach is generating or finding so-called adversarial examples~\cite{goodfellow2014explaining,hendrycks2021natural,athalye2018synthesizing}. Those are characterized as small changes to the input of a neural network that are not \chdeleted{(or only with difficulty)} perceptible by humans but cause the model to flip its decision or predict a different class.  We conceptually follow a similar approach \chadded{to}\chdeleted{as is used for} adversarial samples: we propose to generate inputs to a decision-making model that leads to unexpected results. However, we are explicitly interested in cases where changes are perceptible, but their impact does not align with human expectations. \chdeleted{Thus, we introduce human raters as a comparison to the model's decision to better understand where perception aligns and where the model acts unexpectedly.}

\subsection{\chadded{Model Evaluation and Introspection}}
\chadded{The state of the art for reporting on decision-making performance are global metrics like precision, recall, or F-scores~\cite{powers2020evaluation}.
Beyond these metrics, the field of explainable artificial intelligence (XAI) has proposed different methods for further model introspection~\cite{barredo2020explainable,lipton2016mythos,hohman2018visual,guidotti2018survey,liu2018analyzing,ma2019explaining}. Some researchers divide these approaches into categories such as model explanations, outcome explanations, black box inspection, and transparent design~\cite{lipton2016mythos}, while others distinguish based on whether they focus on explaining single predictions (local) or the model as a whole (global), and whether they are model-specific (applicable to one model or a group of models) or model-agnostic (suitable for any model)~\cite{barredo2020explainable}. Models can be transparent by design or explored with post-hoc explanations~\cite{barredo2020explainable}. Our method provides such post-hoc explanations and is model agnostic, meaning it can be applied to various models. The novelty of our approach is in its inclusion of human elements in the explainability process.}

\section{Method}

We suggest comparing the outputs of a decision-making model on strategically sampled inputs to answers by human raters to better understand the model. Note how both the humans and the generated samples are needed. Without human ratings, we don't gain insights into mismatches in perception, and by using only real-world data, we cannot access the space between clear-cut decisions where we expect those mismatches to be found. Here, we use a generative model to produce samples inspired by the outputs of a classical classification task: true and false positives as well as their negative counterparts. 
\chdeleted{This section gives more details on the steps of this process.}
We illustrate and explore our method for the concrete case of face recognition as an example of a decision-making model. All empirical claims are limited to this use case \chadded{and we have no evidence if our findings generalize to different decision-making models or sample types}. 
However, \chadded{our modular design should allow for adapting it to different use-cases and} we discuss \chadded{possible} extensions and applications of our method in Section \ref{sec:discussion}.

\subsection{Generator}

The core component of our approach is a generative model to provide samples that can serve as input to the model we wish to test. 
For the case of evaluating face recognition, we propose \chadded{using GANs as}\chdeleted{the use of Generative Adversarial Networks (GANs) to generate such samples as}
they have been shown very capable of generating realistic face images~\cite{karras2019style, Karras2020cvpr} and their continuous latent space can be leveraged for targeted manipulations~\cite{shen2020interpreting, wang2021attribute}. 
They can thus be used to produce alterations of a given starting point and samples \textit{between} existing real-world data points for which no ground-truth "true" labeling exists. Note that while we suggest using GANs for face images, other generative approaches (see Section \ref{sec:RW}) are possible as long as they can produce targeted manipulations. In some cases, no model may be required (e.g., if the decision-making model uses only numerical inputs). %

\subsection{Samples}

\begin{figure*}
    \centering
    \vspace{-2mm}
    \begin{subfigure}[t]{.19\linewidth}
         \centering
         \includegraphics[width=.95\linewidth]{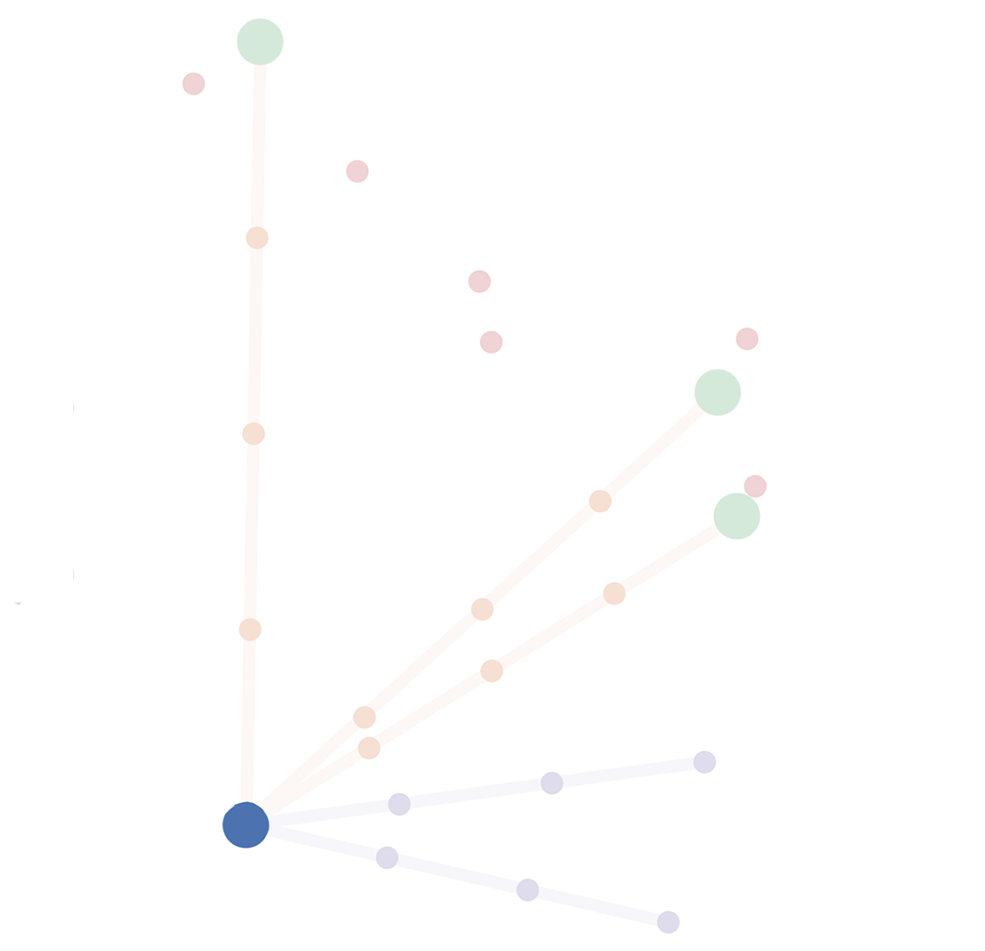}
         \subcaption{Genuine}
         \label{fig:sample_genuine}
    \end{subfigure}
    \hfill
    \begin{subfigure}[t]{.19\linewidth}
         \centering
         \includegraphics[width=.95\linewidth]{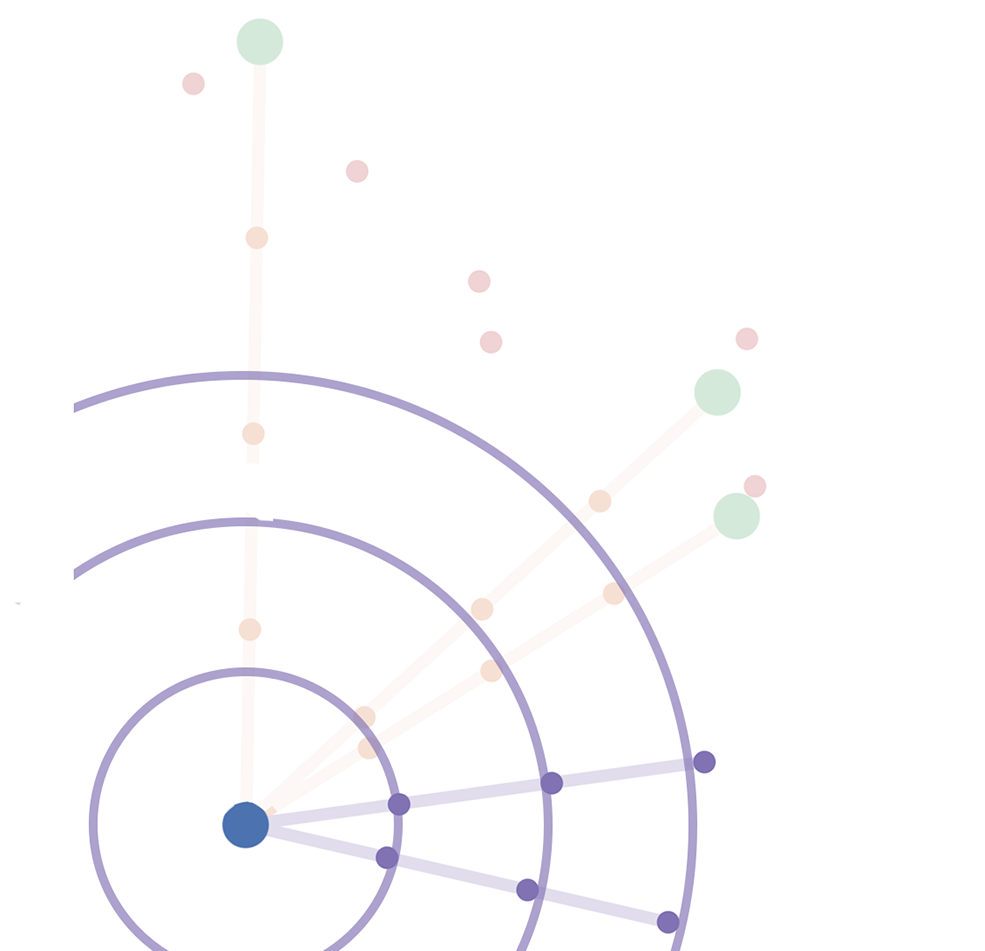}
         \subcaption{Positive}
         \label{fig:sample_positive}
    \end{subfigure}
    \hfill
    \begin{subfigure}[t]{.19\linewidth}
         \centering
         \includegraphics[width=.95\linewidth]{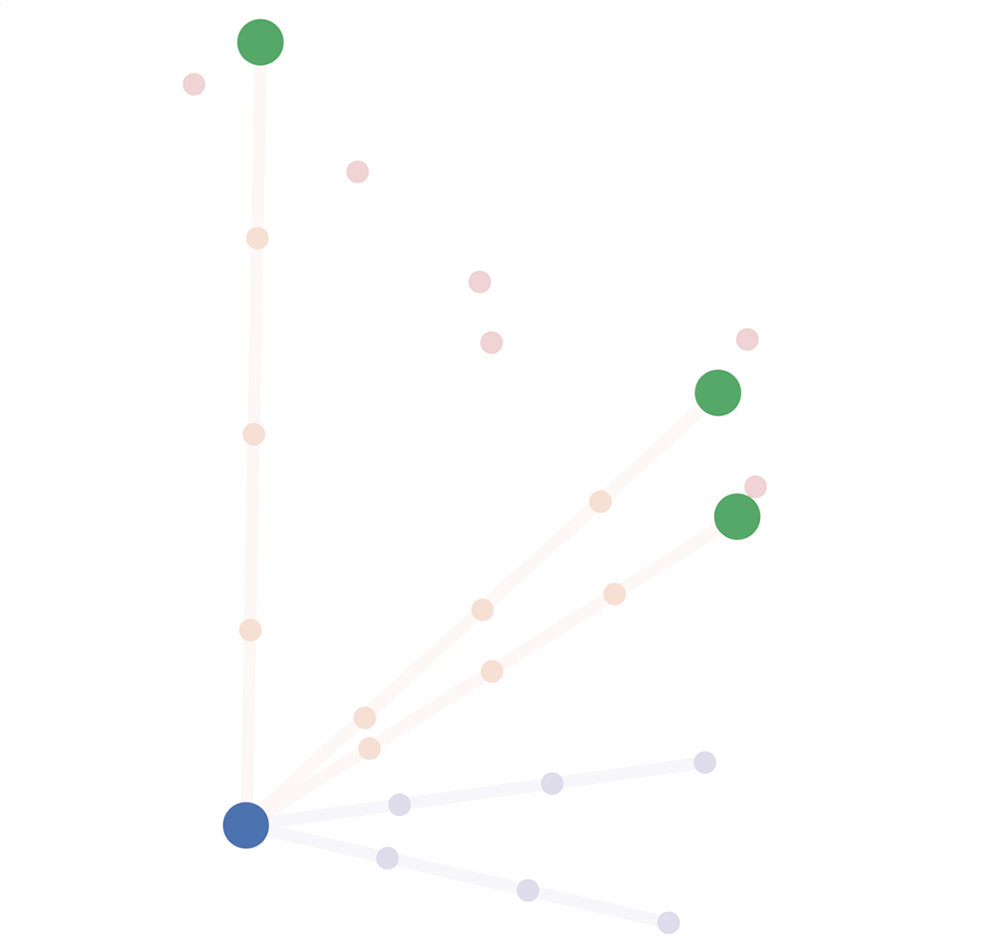}
         \subcaption{Negative}
         \label{fig:sample_negative}
    \end{subfigure}
    \hfill
    \begin{subfigure}[t]{.195\linewidth}
         \centering
         \includegraphics[width=.95\linewidth]{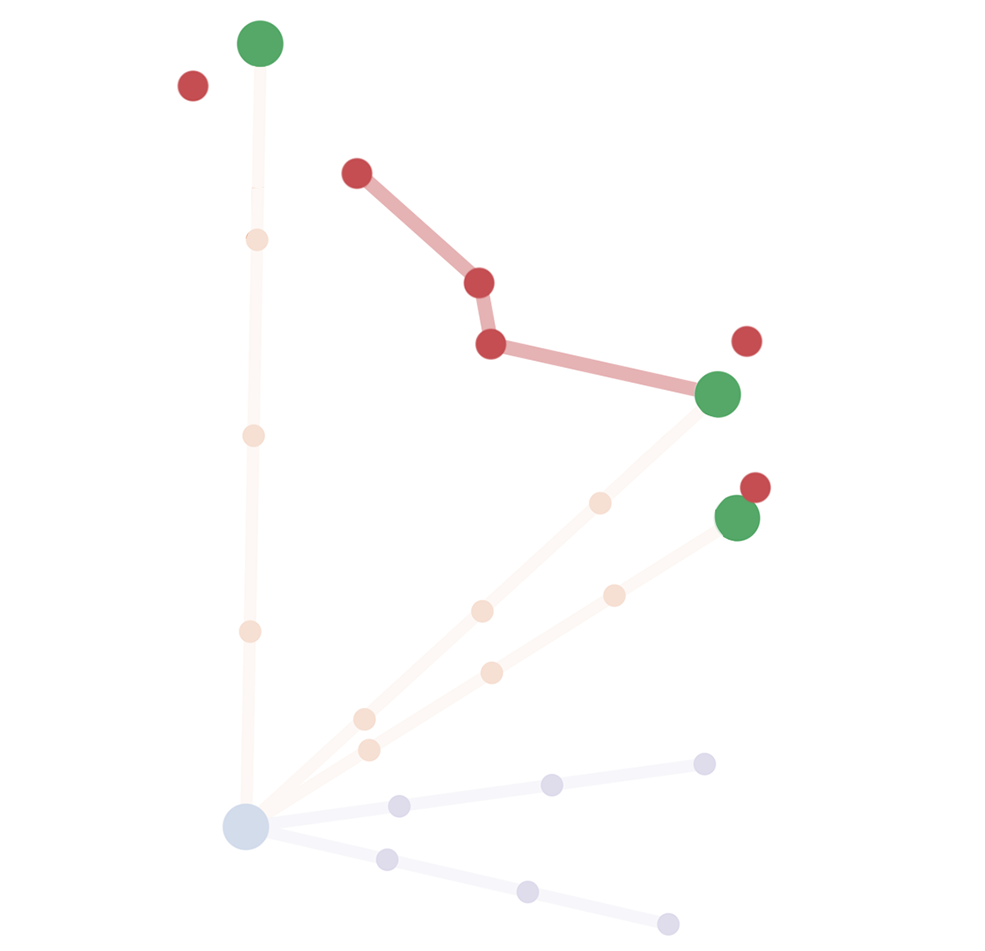}
         \subcaption{Optimized}
         \label{fig:sample_optimized}
    \end{subfigure}
    \hfill
    \begin{subfigure}[t]{.19\linewidth}
         \centering
         \includegraphics[width=.95\linewidth]{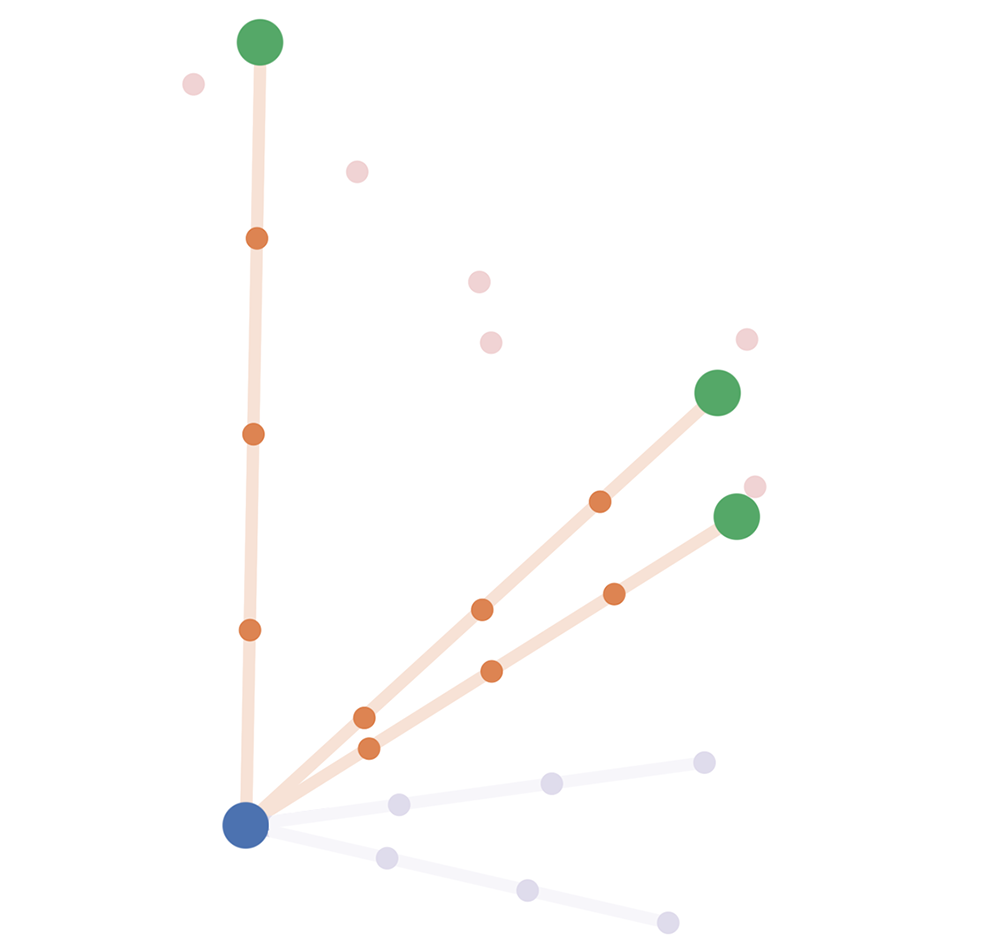}
         \subcaption{Interpolation}
         \label{fig:sample_interpolation}
    \end{subfigure}    
    \caption{Illustration of how our proposed samples are generated (simplified illustration of a latent space). Positive samples are generated by sampling points close the the genuine sample. Negatives are random other points in the latent space. Optimized samples are generated starting at a negative sample and using an optimization function to find samples that are more similar to the genuine sample. Interpolation samples are found as steps on the latent path between the genuine sample and negative samples.}
    \label{fig:samples_illustration}
    \vspace{-3mm}
\end{figure*}

We now illustrate the sets of samples we propose to generate and the rationale behind choosing them. Each sample is always generated in relation to a \textit{base} (i.e., the starting point in the latent space) that will later be used for the comparison (see Section \ref{sec:raters}). For the example of face recognition, this would be the face image to test the decision model on. Figure \ref{fig:samples} illustrates examples for each of the sample types, and Figure \ref{fig:samples_illustration} shows in more detail how they relate to the base image.
\chadded{In this paper, we define samples by their expected decision (e.g., we expect \textit{positive samples} to be rated as similar) and find them by their relation in latent space to avoid making assumptions on possible influencing factors. However, when such assumptions exist (e.g., light conditions affecting recognition), samples could be generated to reflect them instead (see, e.g., \cite{denton2019image}). } 
Note that our generated samples are independent of the decision-making model to be tested. As such, our approach is also suited to explore black-box models as long as they can be queried. 

\subsubsection{Genuine Samples}
Genuine samples (Figure \ref{fig:sample_genuine}) are identical to the base image and, thus, what would be called a true positive in traditional classification tasks, i.e., they preserve the identity of the base image. Note that this is the only possible true positive as it is not necessarily clear to which identity a sample should be attributed. This is the case because we can continuously sample images from the latent space between two identities in contrast to the real world.
We include this sample as a baseline for participants to calibrate their rating of similarity against and as a check to ensure participants pay attention while rating the image pairs. 

\subsubsection{Positive Samples}
We propose positive samples (Figure \ref{fig:sample_positive}) as slight modifications to the base. We find those by sampling a (random) direction in the latent space and taking a small step away from the base in that direction. The assumption behind this approach is that, given a locally stable latent space, this should produce minor alterations to the input and preserve the model's decision on the base (for example, identity for our case of face recognition). 

\subsubsection{Negative Samples}
We choose random samples from the latent space as negative samples (Figure \ref{fig:sample_negative}), i.e., samples that we expect to be attributed to a different identity as the base. This is based on the assumption that the latent space is big enough that randomly generating a sample similar to the base is unlikely. For practical reasons, we propose using other base samples as negative samples. This way, bases can serve as both genuine and negative samples, and their function is only defined relative to their respective base.

\subsubsection{Optimized Samples} 
We propose to use each negative sample as a starting point for a black box optimization algorithm to find samples that maximize the decision-making model's target function (for example, similarity). In contrast to the other samples, this step requires access to either the decision-making model itself or a similar one. In our results (see Section \ref{sec:FR}) we explore if two different models can be used for this (i.e., one to be explored, one for generation). 
Given the different starting points, we assume that the generated samples (Figure \ref{fig:sample_optimized}) should represent local maxima in the latent space (instead of finding the original base sample). Thus, they may very well not be similar to a human observer, even though they are similar according to the optimization function. As such optimized samples fulfill the role of potential false positives. 

\subsubsection{Interpolation Samples} 
To better understand where the model's decision between two (base) samples changes, we introduce interpolation samples (Figure \ref{fig:sample_interpolation}). Those are generated by following the latent vector between the base and each negative sample. 
Generating candidates for false negatives through optimization would require human ratings as a target function. As those are not available at generation time, interpolation samples are our best attempt at provoking this type of misclassification. 

\subsection{(Human) Raters}\label{sec:raters}
The final component of our approach is having the generated samples rated by humans as a baseline to compare against the model's decisions (remember that for the samples we propose, no ground truth is available; so this step is necessary). This is based on the premise that the decision-making model is supposed to decide similarly to a human or at least in a way that aligns with human intuition. To reflect this, we propose that human raters judge the model's target function and their expected decision. For the example of a face recognition model, this maps to a similarity score for the presented faces and the decision whether two images show the same person. Depending on the use case, it may be possible to have the samples rated by a different model instead of humans if only differences between models should be explored with no focus on whether they adhere to human perception. 

\section{Experiments}

We now illustrate how we implemented the approach described above for the example of face recognition. We explain how we generated the samples and implemented the comparison task for human raters. \chadded{With this experiment, we explore the robustness of our proposed method and uncover potential misalignment between human raters and face recognition models.} Note that latent spaces allow for many potential comparisons, and our choice can only capture a fraction of them. The choices of both samples and parameters presented here reflect our best attempt at striking a balance between many potential comparisons, sufficiently many samples to observe trends, and a number of comparisons that participants can realistically make.

\subsection{Dataset Generation}\label{sec:dataset}

We generate all samples using StyleGAN2~\cite{Karras2020cvpr}. Base images (and consequently negative and genuine images) are sampled as random seeds from the latent space. The number of samples scales with the number of base images, as, e.g., interpolation samples are generated between all available bases. Thus, we decided on a batched approach with four base images per batch. For each base, we generate one genuine sample (i.e., a copy). We chose two random directions and generated three positive samples (at distances 0.2, 0.4, and 0.6) for each of them (6 total). As negative samples, we include the remaining three base images. Furthermore, we generate interpolation samples at 25\%, 50\%, and 75\% of the distance between the base and each negative sample (9 total). Finally, we use the Covariance Matrix Adaptation Evolution Strategy (short CMA-ES)\footnote{\url{https://pypi.org/project/cma/}\accessed} as a black box optimizer to find optimized samples. This approach empirically finds a gradient by sampling points around a given starting location and optimizing a given score function. We encoded images as a 512-dimensional vector of their latent representation and used a Python face recognition algorithm\footnote{\url{https://pypi.org/project/face-recognition/}\accessed} as the optimization function. We ran 100 iterations with a truncation factor of 0.3 (i.e., we stopped optimization when reaching this distance score) using $\sigma$ = 1. We ran one optimization starting from each of the negative samples. We chose the first results that achieved a recognition distance below 0.3, 0.4, and 0.5 respectively\footnote{Due to an error in the implementation, the images chosen for distances 0.4 and 0.5 included samples with worse ratings} as optimized samples (9 in total). The best sample was chosen if the optimization did not reach this score. For reference: the suggested default recognition distance for deciding on identity in the used Python library is 0.6, so all of those samples would be accepted.
Overall, this approach yielded 112 samples in each batch (28 samples for each base image).

\subsection{Procedure}

\begin{figure*}
    \centering
    \includegraphics[width=.65\linewidth]{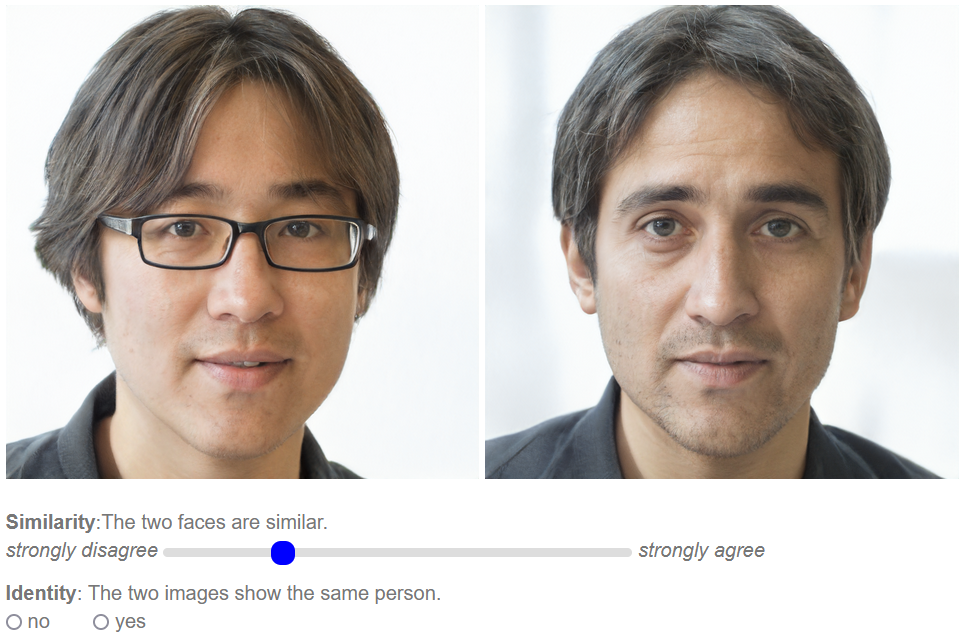}
    \caption{Screenshot of the main task in our online study. Participants were presented with two images (a base image on the left and a sample generated with our approach on the right) and were asked to rate their similarity and if the images showed the same person.}
    \label{fig:study}
\end{figure*}

To collect human ratings on our samples, we designed an online survey. First, participants would be informed about the purpose of the study and had to consent to their data being collected and analyzed. Next, we collected basic demographic data before participants got to the main task. Here, participants were repeatedly presented with an image pair where one was always a base image and the other was one of the samples described in Section \ref{sec:dataset}. Each participant rated 112 image pairs \chadded{with their order being randomized both between and within participants.}\chdeleted{in randomized order.} For each image pair, we asked participants for their perception of the similarity of the two faces and a binary decision if they believed they showed the same person (see Figure \ref{fig:study} for an example of such a choice). \chadded{We kept the overall study duration short to avoid the repetitive nature of the task influencing our results.}
We concluded by asking participants for their strategy in rating both similarity and identity. 

\subsection{Participants and Recruitment}

We recruited 100 participants (50 female, 48 male, 1 non-binary, 1 no answer) with a mean age of 27.6 (STD=7.8, range: 19-61) using Prolific\footnote{\url{https://prolific.co}\accessed}. People from all continents contributed to our dataset. The study took about 20 minutes and was compensated with £2.25. The study was approved by our institute's ethics commission under Nr. EK-MIS-2023-204. 

\section{Results}

In this section, we describe the collected dataset\chadded{ and}\chdeleted{,} verify the effectiveness \chadded{and robustness} of our sampling methods\chadded{. We conduct a correlation analysis to assess the alignment of human ratings and model scores and propose disagreement scores to find visual examples for the cases where ratings differ the most.}\chdeleted{, and demonstrate how our approach can be used to gain insights into the decision-making model. } 

\subsection{Dataset overview}

\begin{table*}[tb]
    \centering
    \caption{\chadded{Mean similarity (sim) and identity (id) ratings of our participants and distance scores of different face recognition models (Dlib, VGG: VGG-Face, FN: FaceNet512, OF: OpenFace) and distance metrics (lpips, latent) for the tested sample types.}\chdeleted{Descriptive overview of our dataset.} Arrows indicate the direction of images being perceived \chadded{more strongly as the same person or} as more similar. In brackets, we indicate the acceptance rates (0: all samples were rated as different identities, 1: all samples were rated as the same identity) of \chdeleted{human raters and} face recognition models \chdeleted{(sim: rated similarity, VGG: VGG-Face, FN: FaceNet512, OF: OpenFace)} based on their default recognition thresholds (Dlib: 0.6, VGG: 0.86, FN: 1.04, OF: 0.55).}
    \label{tab:desc}
    \begin{tabular}{lr>{\rowmac}r|r@{\hskip 0.1cm}>{\rowmac}lr@{\hskip 0.1cm}>{\rowmac}lr@{\hskip 0.1cm}>{\rowmac}lr@{\hskip 0.1cm}>{\rowmac}l|rr}
    \toprule
    
    \setrow{\itshape}& \multicolumn{2}{c}{\textbf{perceived}}&\multicolumn{8}{c}{\textbf{face recognition models}}&\multicolumn{2}{c}{\textbf{distance metrics}}\\
    
    Sample type & sim $\uparrow$ & id $\uparrow$ & \multicolumn{2}{c}{Dlib $\downarrow$ \chadded{($\uparrow$)}} & \multicolumn{2}{c}{VGG $\downarrow$ \chadded{($\uparrow$)}}& \multicolumn{2}{c}{FN $\downarrow$ \chadded{($\uparrow$)}}  & \multicolumn{2}{c}{OF $\downarrow$ \chadded{($\uparrow$)}} & lpips $\downarrow$ & latent $\downarrow$ \\
    
    \midrule

    \textbf{Genuine} & 97.75 & 0.98 & 0.00 & (1.00) & 0.00 &(1.00) & 0.00 & (1.00) & 0.00 & (1.00) & 0.00 & 0.00 \\
    \textbf{Interpolation} & 39.97 & 0.25 & 0.61 & (0.43) &0.92 & (0.40) & 1.12 & (0.32) & 0.85 & (0.17) & 0.52 & 67.87 \\
    \hphantom{***}-- dist 25\% & 73.30 & 0.59 & 0.42 & (0.88) & 0.76 &(0.66) & 0.90 & (0.64) & 0.74 & (0.33) & 0.37 & 33.94 \\
    \hphantom{***}-- dist 50\% & 31.35 & 0.13 & 0.67 & (0.35) & 0.96 &(0.33) & 1.18 & (0.23) & 0.86 & (0.10) & 0.56 & 67.87 \\
    \hphantom{***}-- dist 75\% & 15.26 & 0.02 & 0.76 & (0.07) & 1.05 &(0.22) & 1.29 & (0.07) & 0.94 & (0.07) & 0.64 & 101.81 \\
    \textbf{Negative} & 11.34 & 0.02 & 0.80 & (0.02) & 1.08 &(0.18) & 1.33 & (0.02) & 0.93 & (0.07) & 0.67 & 135.75 \\
    \textbf{Optimized} & 35.36 & 0.16 & 0.50 & (0.68) & 0.89 &(0.48) & 1.08 & (0.41) & 0.86 & (0.12) & 0.60 & 227.16 \\
    \hphantom{***}-- dist 0.3 & 51.67 & 0.31 & 0.31 & (0.99) & 0.76 &(0.68) & 0.92 & (0.70) & 0.80 & (0.22) & 0.55 & 279.99 \\
    \hphantom{***}-- dist 0.4 & 32.49 & 0.11 & 0.53 & (0.80) & 0.88 &(0.53) & 1.07 & (0.46) & 0.85 & (0.11) & 0.61 & 217.45 \\
    \hphantom{***}-- dist 0.5 & 21.91 & 0.05 & 0.66 & (0.25) & 1.01 &(0.22) & 1.26 & (0.07) & 0.95 & (0.03) & 0.64 & 184.05 \\
    \textbf{Positive} & 72.32 & 0.57 & 0.41 & (0.86) & 0.70 &(0.67) & 0.87 & (0.65) & 0.68 & (0.40) & 0.34 & 38.11 \\
    \hphantom{***}-- dist 0.2 & 88.46 & 0.84 & 0.29 & (0.99) & 0.55 &(0.79) & 0.68 & (0.81) & 0.57 & (0.61) & 0.24 & 19.06 \\
    \hphantom{***}-- dist 0.4 & 70.97 & 0.53 & 0.43 & (0.88) & 0.72 &(0.69) & 0.89 & (0.65) & 0.69 & (0.36) & 0.36 & 38.11 \\
    \hphantom{***}-- dist 0.5 & 57.52 & 0.34 & 0.52 & (0.72) & 0.84 &(0.54) & 1.03 & (0.50) & 0.78 & (0.24) & 0.43 & 57.17 \\
    \midrule
    \textbf{all} & 44.42 & 0.29 & 0.53 & (0.58) & 0.85 &(0.48) & 1.04 & (0.41) & 0.80 & (0.22) & 0.51 & 117.54 \\
    \bottomrule
    \end{tabular}
\end{table*} 

Our dataset consists of 1,120 image pairs (10 batches of 112 image pairs each) that were rated by 10 participants each, resulting in a total of 11,200 ratings of similarity \chadded{(0: not similar to 100: very similar)} and identity \chadded{(fraction of participants rating as identical)}. In addition, we post-hoc calculated distance scores for four common state-of-the-art face recognition algorithms using the DeepFace library by \citet{serengil2020lightface}, as well as latent distance (based on the distance of embeddings in the StyleGAN2 latent space) and perceptual distance (lpips)~\cite{zhang2018perceptual}. For the sake of brevity, we only compare against Dlib face distance when making comparisons to a face recognition model (unless otherwise stated). An overview of our dataset grouped by type of sample is given in Table \ref{tab:desc}.

\subsection{\chadded{Robustness of Sample Generation based on Human Ratings}\chdeleted{Human ratings with respect to sample types}}\label{sec:discussion_optimized}

\begin{figure*}[tb]
    \centering
    \begin{subfigure}[t]{.49\linewidth}
         \centering
         \includegraphics[width=\linewidth]{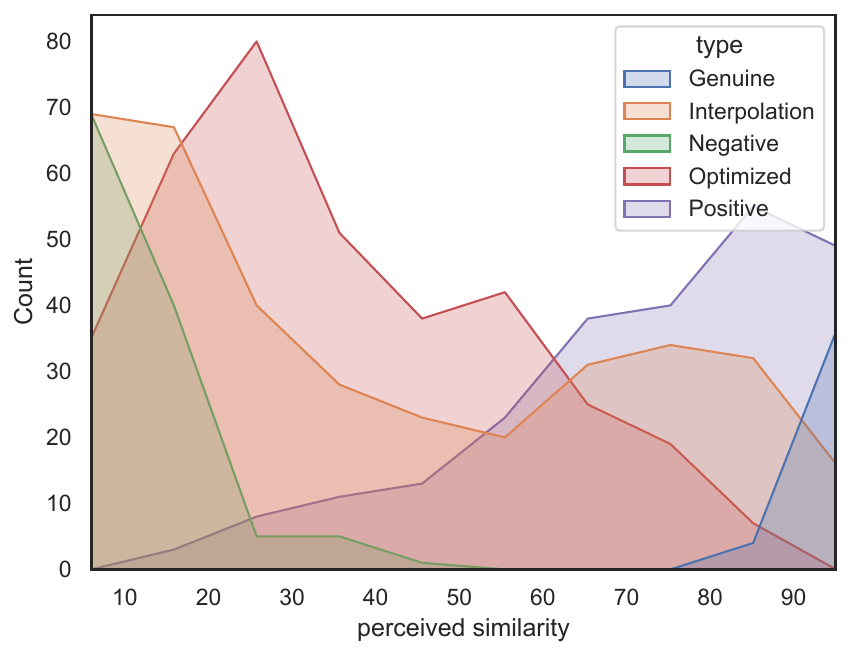}
         \subcaption{\chadded{Histogram}\chdeleted{Distribution} of ratings of perceived similarity for the different proposed sample types \chadded{(lines instead of bars for easier readability}.}
         \label{fig:type_overview}
    \end{subfigure}
    \hfill
    \begin{subfigure}[t]{.49\linewidth}
         \centering
         \includegraphics[width=\linewidth]{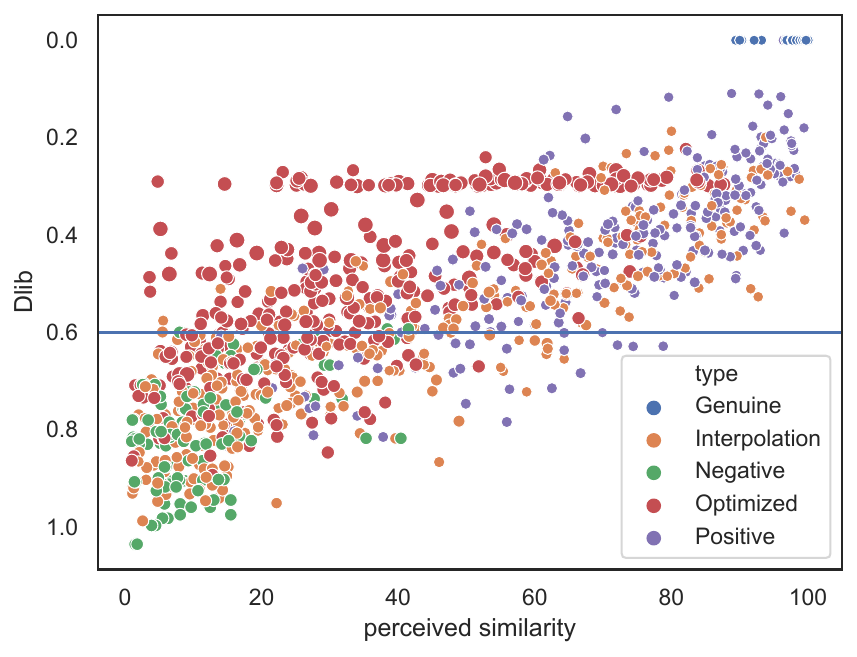}
         \subcaption{\chadded{Scatter plot of} perceived similarity compared to the rating of a face recognition model (Dlib, line: default decision threshold) by \chadded{sample} type. \chadded{Marker size indicates latent distance.}}
         \label{fig:type_FR}
    \end{subfigure}
    \caption{Distribution of perceived similarity based on the type of sample (left) and in comparison to the ratings of a face recognition model (right). \chdeleted{Marker size indicates latent distance.} }
    \label{fig:type}
\end{figure*}

As a first step, we validate the \chadded{robustness}\chdeleted{success} of our approach. An overview of the distribution of \chadded{human} similarity ratings by sample type is given in Figure \ref{fig:type_overview}. \chadded{Here, we report on mean similarity and identity ratings as presented in Table \ref{tab:desc}}. As expected, genuine samples were rated as both similar (\chadded{Mn=}97.75) and the same person (\chadded{Mn=}0.98). The rating of interpolation samples strongly depended on the interpolation distance. Samples at 25\% of the distance away from the genuine samples were rated highly similar and often still as the same person. Similarity decreased with more distance, as did the identity rating until at 75\%, they were similar to the ratings for negative samples (similarity: \chadded{Mn=}11.34, id: \chadded{Mn=}0.02). As designed, optimized samples were often rated as fairly similar (\chadded{Mn=}35.36) but not as having the same identity (\chadded{Mn=}0.16). Finally, positive samples were rated as similar (\chadded{Mn=}88.46) and the same person (\chadded{Mn=}0.84) for close distances as intended. However, this effect strongly decreased for larger distances. \chadded{While ratings were generally not unanimous (see Figure \ref{fig:type_overview}), their distributions were clearly distinguishable and in line with our expectations, showing that our approach was effective and robust in provoking the desired participant responses.}

\subsection{\chadded{Alignment of Human Ratings with}\chdeleted{Comparing (human ratings to)} Face Recognition Models and Distance Metrics}\label{sec:FR}

\begin{table*}[b]
    \centering
    \caption{\chadded{Pearson} correlation of \chadded{human-rated similarity, face recognition scores and distance metrics}\chdeleted{different distance metrics} with human-rated identity. \chadded{Values close to 1/-1 indicate strong alignment with human perception of identity} We omit genuine samples as distance scores for them were mostly constant (see Table \ref{tab:desc}) and correlations thus are invalid.}\label{tab:fr_cor}
    \begin{tabular}{lr|rrrr|rr}
    \toprule
    & \multicolumn{1}{c}{\textbf{perceived}}&\multicolumn{4}{c}{\textbf{face recognition models}}&\multicolumn{2}{c}{\textbf{distance metrics}}\\
    Sample type & similarity \chadded{$\uparrow$} & Dlib \chadded{$\downarrow$} & VGG-Face \chadded{$\downarrow$} & Facenet512 \chadded{$\downarrow$} & OpenFace \chadded{$\downarrow$} & lpips \chadded{$\downarrow$} & latent \chadded{$\downarrow$} \\
    \midrule

    \textbf{Interpolation} & 0.901** & -0.796** & -0.528** & -0.668** & -0.345** & -0.738** & -0.719** \\
    \textbf{Negative} & 0.637** &  & -0.334** & -0.21*\hphantom{*} & &  &  \\
    \textbf{Optimized} & 0.846** & -0.589** & -0.313** & -0.38** & -0.204** & -0.494** & 0.347** \\
    \textbf{Positive} & 0.838** & -0.787** & -0.556** & -0.614** & -0.389** & -0.71** & -0.634** \\
    \midrule
    \textbf{all} & 0.907** & -0.765** & -0.632** & -0.712** & -0.512** & -0.814** & -0.453** \\
    \bottomrule
    \multicolumn{8}{c}{\footnotesize{*: p < .05, **: p < .001, empty cells: not significant}}
    \end{tabular}
\end{table*}

\begin{figure*}[tb]
    \centering
    \includegraphics[width=0.48\linewidth]{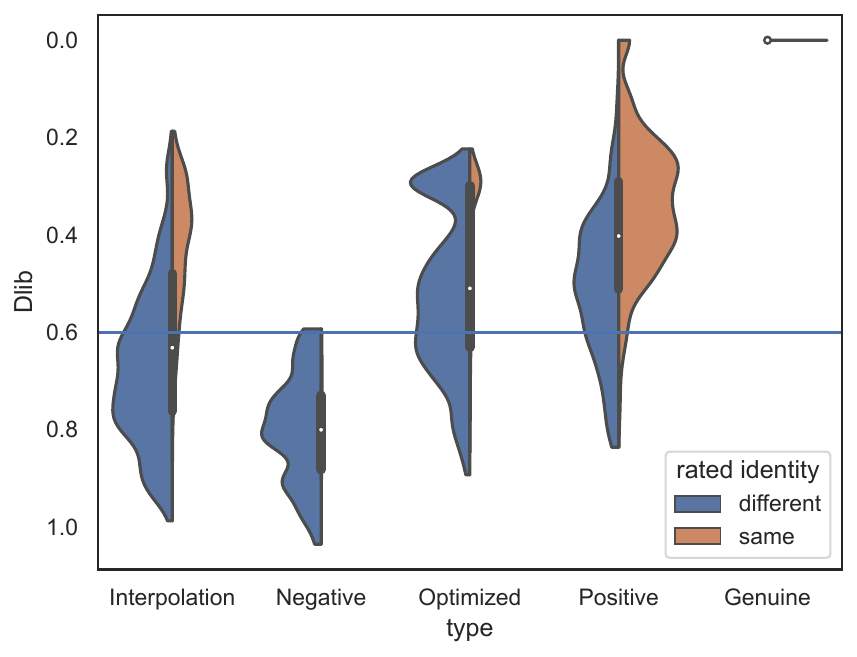}
    \includegraphics[width=0.48\linewidth]{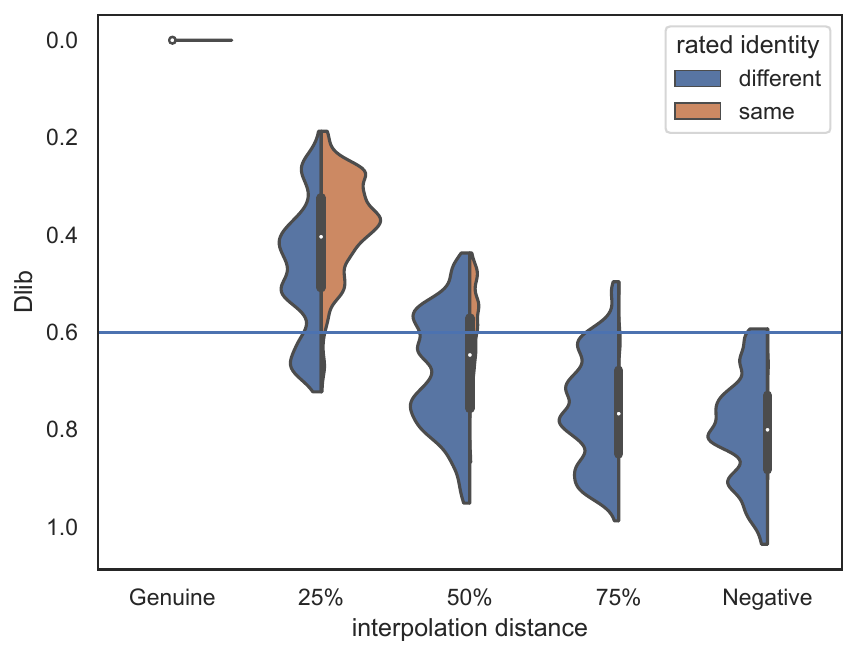} 
    \caption{\chadded{Violin plots showing the distribution of face recognition scores for different sample types split by the human identity rating (blue: rated as different by humans, orange: rated as the same person). The width of each violin is proportional to the number of observations. The left figure shows ratings and scores for all sample types, and the right figure gives more details on the different levels of interpolation.}\chdeleted{Rated identity with respect to the different types of samples and the associated face recognition score (left). Influence of the interpolation on rated identity (right).} The blue line indicates the default distance for the face recognition model to accept a face. }
    \label{fig:id}
\end{figure*}

\chadded{Next, we assessed the alignment of human ratings with model decisions.} Figure \ref{fig:type_FR} shows how the perceived similarity of our human raters \chadded{followed a similar trend as}\chdeleted{was related to} the rating by a face recognition algorithm. \chdeleted{, showing that both ratings are generally correlated.} Points farther away in the latent space were generally rated less similar, which is in line with related work~\cite{shimizu2022human}. However, our optimized samples break both trends, being rated less similar by humans but more similar by the algorithm while also farther away in latent space\chadded{, hinting at a misalignment}. We generated the optimized samples with Dlib distance as the target function. Identification results (see Table \ref{tab:desc}) show that they were also most effective at being recognized by this very approach. However, the acceptance rate of Dlib was higher overall than the other models, so we can draw no conclusions if generating optimized samples with a different model than the one to be tested is an effective approach in general. 

We conducted \chadded{a Pearson correlation of human-rated identity with rated similarity, face recognition scores, and distance metrics}\chdeleted{ a correlation analysis of perceived human identity }(see Table \ref{tab:fr_cor}) to better understand those effects. \chadded{Human ratings of similarity and identity were overall strongly correlated. Only negative samples were rated as different identities but sometimes perceived as similar (see Table \ref{tab:desc}). However, such coincidental similarity is to be expected due to our approach of sampling random other images as negative samples.} We found all \chadded{face recognition models and distance metrics}\chdeleted{measures} to correlate most with rated identity for interpolation and positive samples\footnote{\chadded{Correlation values are negative here, as we compare a similarity to a distance metric.}}. Correlation for optimized samples was weaker overall, showing that they fulfilled their purpose to resemble samples whose identification by a face recognition algorithm is not well aligned with human perception. OpenFace showed a weaker correlation than the others and also identified fewer samples as the same person (see Table \ref{tab:desc}), hinting at a stricter model overall. Latent and perceptual distance (lpips)~\cite{zhang2018perceptual} were surprisingly well aligned with identity ratings by our participants. 

\chadded{The violin plot in} Figure \ref{fig:id} gives more detailed insights into the distribution of samples rated as either the same or a different person. Ideally, we would expect to only see \chadded{samples rated as identical}\chdeleted{identical samples} (orange) above the model's threshold and samples rated as different persons (blue) below. Both genuine and negative samples follow this trend, while optimized and positive samples have large areas where their ratings show a mismatch between model and humans. For interpolation samples, mismatches are mainly concentrated on the first interpolation step at 25\%. 

\subsection{Finding Disagreement \chadded{between Human Raters and Models}}

\begin{figure*}
    \centering
    \begin{subfigure}[t]{.49\linewidth}
         \centering
         \includegraphics[width=\linewidth]{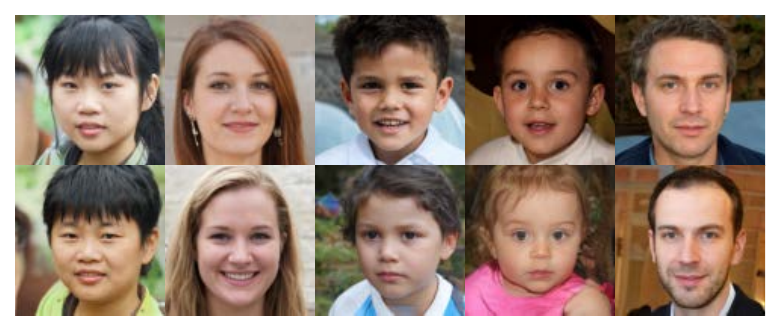}
         \subcaption{Samples with the largest disagreement between human raters and Dlib recognition score (\textit{"False positive"} samples). }
         \label{fig:dis_dlib}
    \end{subfigure}
    \hfill
    \begin{subfigure}[t]{.49\linewidth}
         \centering
         \includegraphics[width=\linewidth]{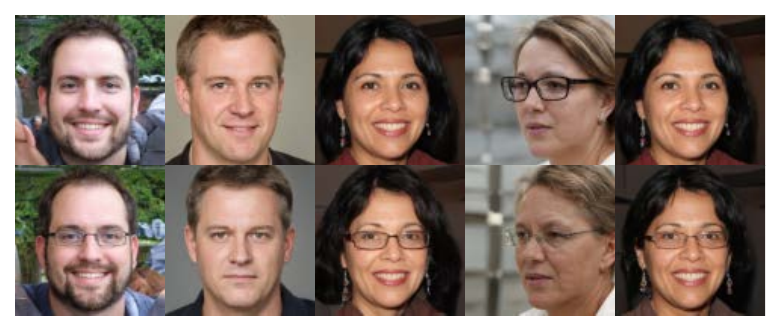}
         \subcaption{Samples with the largest disagreement between human raters and Dlib recognition score (\textit{"False negative"} samples).}
         \label{fig:dis_perc}
    \end{subfigure}
    \begin{subfigure}[t]{.49\linewidth}
         \centering
         \includegraphics[width=\linewidth]{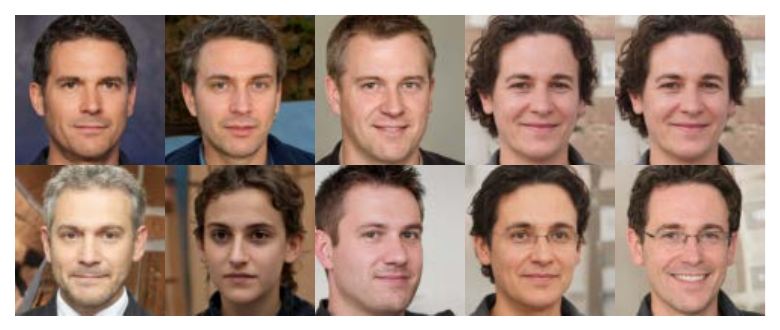}
         \subcaption{Samples with the largest disagreement between participants about \textit{perceived similarity}}
         \label{fig:dis_within_sim}
    \end{subfigure}
    \hfill
    \begin{subfigure}[t]{.49\linewidth}
         \centering
         \includegraphics[width=\linewidth]{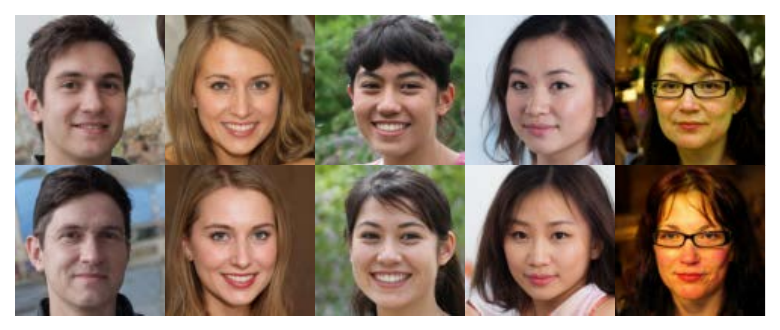}
         \subcaption{Samples with the largest disagreement between participants about \textit{perceived identity}}
         \label{fig:dis_within_id}
    \end{subfigure}    
    \caption{Samples from our dataset with the biggest disagreement between model and participants (\ref{fig:dis_dlib}, \ref{fig:dis_perc}) and between participants themselves (\ref{fig:dis_within_sim}, \ref{fig:dis_within_id}). The top row in each figure contains base images and the bottom row contains generated alterations.}
    \label{fig:diagreement}
\end{figure*}

One of the goals of our approach was to find samples that lead to a mismatch between human raters and a decision-making model or are generally challenging to decide. To find such cases, we calculated a disagreement score between human-rated identity and the (inverted) Dlib distance score. We do not use an absolute value, as both directions are interesting and map to likely candidates for false positives (face recognition rating similarity higher) and false negatives (humans rating similarity higher). To find disagreements between participants, we calculate the standard deviation of their ratings of similarity and identity. Figure \ref{fig:diagreement} shows the samples with the largest disagreement with respect to the described measures. 
We observe that the candidates for false positives (Figure \ref{fig:dis_dlib}) are mainly children and were generated by optimized samples, hinting at a weakness in the assessed decision-making model to correctly judge the similarity of children. Potential false negatives (Figure \ref{fig:dis_perc}) were mainly generated through negative and interpolation samples. They have in common, that they differ in meaningful ways like age, pose, or accessories. 

\subsection{Determining Similarity}
We asked participants for \chadded{the} strategies \chadded{they used} to determine the similarity and identity of the given image pairs. Most participants stated that they compared facial features and decided based on faces looking similar or following their intuition. The main features participants looked at were eyes and eye color, hair (color), the nose, and the mouth area. When judging identity, participants also particularly focused on identifying details, for example, P3: ``I was looking for particular facial features (for example, dimples, teeth, wrinkles, nose shape, etc.)''. They also tried to ignore parts of the image that did not contribute to identity: ``If the majority of faci[a]l features were identical and only the hair or clothe[s] changed I assumed that the images showed the same person.''(P5). 

In addition to our participants' feedback, we investigated influencing factors on perceived similarity computationally. We used a facial feature prediction model on all generated samples to generate a vector of 37 features. We calculated the distance of those vectors between each sample and its respective base. We trained a random forest regressor to predict the collected perceived similarity from those vectors, achieving a $R^2$ score of 0.45. We calculated permutation importance for this model to find the features with the greatest impact on the decision. The most impactful features were related to age (bags under eyes), face shape (oval face, high cheekbones), and styles (hair and makeup).

\section{Discussion}\label{sec:discussion}

In this work, we studied one example case for our proposed method to explore decision-making models. Here, we reflect on our insights from this test, the limitations of our approach, and further target groups and applications to explore.

\subsection{Insights}

We found that human perception overall aligned with the outputs of our test face recognition model. We found a strong correlation between human identity ratings and perceptual distance (lpips)~\cite{zhang2018perceptual}, indicating it may be well suited as an approximation for face recognition. However, our generated samples were also successful in uncovering misalignments. Optimized samples were very successful at generating potential cases of false positives. Similarly, positive and interpolation samples led to cases where the decision-making model indicated less similarity compared to human raters. Visual inspection of the samples with the largest disagreement suggests that the tested model struggled with distinguishing children and was affected by changes like age, pose, or accessories that did not affect the identity for human raters. \chadded{This knowledge can be leveraged to improve the models or inform end-users that such} \chdeleted{This has different implications for different groups. As users of the model, one has to be aware that} small everyday changes can largely impact recognition performance. \chdeleted{As a developer of such a system, this gives starting points for what to improve. }For our method, the strong disagreement for samples with meaningful but minor edits implies that specifically sampling for such differences may be a good way of finding potential false negative samples and should be considered as an addition to our proposed \chadded{sampling strategy}\chdeleted{samples}.

\vspace{-2mm}
\subsection{\chadded{Reflections on the Method}\chdeleted{Considerations}}
Using a generative model to explore a decision-making model can introduce new biases or hide existing shortcomings of the tested models if biases align. While this cannot completely be avoided, we believe that introducing human raters can uncover some of those effects. At the same time, the inclusion of human raters also limits our approach to tasks that actually can be reliably rated by humans. This includes in particular decisions that people already make in their daily lives, like recognizing others by their face, voice, or the way they walk. However, comparing something like fingerprints or making a diagnosis on medical data is uncommon or requires experts, making it less suited for our method. \chadded{While we had a diverse set of raters and tested face recognition algorithms, our results may not generalize to other populations and models. Future work could further investigate such influencing factors.} Finally, we used StyleGAN2~\cite{karras2019style} in this work. However, our approach should be compatible with other GAN variants and potentially also other generative methods, as long as their latent space is locally stable and can be navigated. 

\vspace{-2mm}
\subsection{\chadded{Implications for (Security) User Interfaces}}

Our analysis has shown that our proposed sample sets reliably produce potential mismatches between human perception and model decisions. \chadded{This opens up opportunities for interface design. As an example,}\chdeleted{. Thus,} our approach could be used to \chadded{make model performance graspable}\chdeleted{illustrate model performance} for users \chadded{by illustrating how someone would look like that could unlock their device or how changes in appearance affect the decision. No external human ratings will be needed, as users can directly assess alignment with their perception.}\chdeleted{ In that case, no explicit rating will be needed. The user can make the comparison of their own perception to the given system response by themself to then decide to, e.g.,} \chadded{Model alignment with human perception (based on the results in Section \ref{sec:FR}) can be shown in addition to global performance metrics to help users to} adjust decision thresholds, avoid failure cases, \chadded{choose a different model} or -- more generally -- make an informed decision if and how to use \chadded{it}\chdeleted{the model}.
\chdeleted{Finally, our approach can also be used to explore a model as a whole, as we did in our analysis, giving potential starting points for further detailed analysis, as just described.}

\vspace{-2mm}
\subsection{\chadded{Extension to other Use Cases}\chdeleted{Further Application Areas and Target Groups}}

We used our proposed method for face recognition in this paper. However, we \chadded{chose a modular design that allows for replacing components to facilitate adaptations and thus} expect it to be applicable to other models as well. The key prerequisites we see are: (1) humans can \chadded{make the same decisions as the model (e.g. assess identity or similarity)}\chdeleted{rate the model's decisions}, and (2) generative models exist for the type of input (e.g., voice recognition~\cite{ren2020fastspeech} or gait and gesture recognition~\cite{zhang2023generating}). In addition, our approach also generates a labeled dataset of synthetic samples that can be used to improve the model.

\section{Conclusion}

In this paper, we proposed leveraging generative models to strategically produce alterations (positive, negative, interpolation, and optimized) to the input of a decision-making model and compare its ratings to human ratings. We collected a dataset of 11,200 ratings of similarity and identity for pairs of face images and compared them to the output of a face recognition model, providing insights into how the perception of humans and the algorithm differs. We hope, that our approach can be leveraged by developers of decision-making models to explore and improve weaknesses in their models and support the design of user interfaces for individuals to better understand the models they interact with in their daily lives and empower them to make informed decisions on if and how to use them. 

\section{Resources}

As part of our contribution, we provide the dataset generated in our study. This includes all sample images, their latent codes, and the identity and similarity ratings by our participants and the recognition models. The dataset will be available at \url{https://www.unibw.de/usable-security-and-privacy/research/datasets}.

\begin{acks}
This project is funded by the Deutsche Forschungsgemeinschaft
(DFG, German Research Foundation) – 425869382 and is part of
Priority Program SPP2199 Scalable Interaction Paradigms for Pervasive Computing Environments.

\end{acks}

\bibliographystyle{ACM-Reference-Format}
\bibliography{bibliography}


\begin{thebibliography}{43}


\ifx \showCODEN    \undefined \def \showCODEN     #1{\unskip}     \fi
\ifx \showDOI      \undefined \def \showDOI       #1{#1}\fi
\ifx \showISBNx    \undefined \def \showISBNx     #1{\unskip}     \fi
\ifx \showISBNxiii \undefined \def \showISBNxiii  #1{\unskip}     \fi
\ifx \showISSN     \undefined \def \showISSN      #1{\unskip}     \fi
\ifx \showLCCN     \undefined \def \showLCCN      #1{\unskip}     \fi
\ifx \shownote     \undefined \def \shownote      #1{#1}          \fi
\ifx \showarticletitle \undefined \def \showarticletitle #1{#1}   \fi
\ifx \showURL      \undefined \def \showURL       {\relax}        \fi
\providecommand\bibfield[2]{#2}
\providecommand\bibinfo[2]{#2}
\providecommand\natexlab[1]{#1}
\providecommand\showeprint[2][]{arXiv:#2}

\bibitem[Athalye et~al\mbox{.}(2018)]%
        {athalye2018synthesizing}
\bibfield{author}{\bibinfo{person}{Anish Athalye}, \bibinfo{person}{Logan Engstrom}, \bibinfo{person}{Andrew Ilyas}, {and} \bibinfo{person}{Kevin Kwok}.} \bibinfo{year}{2018}\natexlab{}.
\newblock \showarticletitle{Synthesizing robust adversarial examples}. In \bibinfo{booktitle}{\emph{International conference on machine learning}}. PMLR, \bibinfo{pages}{284--293}.
\newblock


\bibitem[Balakrishnan et~al\mbox{.}(2021)]%
        {balakrishnan2021towards}
\bibfield{author}{\bibinfo{person}{Guha Balakrishnan}, \bibinfo{person}{Yuanjun Xiong}, \bibinfo{person}{Wei Xia}, {and} \bibinfo{person}{Pietro Perona}.} \bibinfo{year}{2021}\natexlab{}.
\newblock \showarticletitle{Towards causal benchmarking of bias in face analysis algorithms}.
\newblock In \bibinfo{booktitle}{\emph{Deep Learning-Based Face Analytics}}. \bibinfo{publisher}{Springer}, \bibinfo{pages}{327--359}.
\newblock


\bibitem[{Barredo Arrieta} et~al\mbox{.}(2020)]%
        {barredo2020explainable}
\bibfield{author}{\bibinfo{person}{Alejandro {Barredo Arrieta}}, \bibinfo{person}{Natalia Díaz-Rodríguez}, \bibinfo{person}{Javier {Del Ser}}, \bibinfo{person}{Adrien Bennetot}, \bibinfo{person}{Siham Tabik}, \bibinfo{person}{Alberto Barbado}, \bibinfo{person}{Salvador Garcia}, \bibinfo{person}{Sergio Gil-Lopez}, \bibinfo{person}{Daniel Molina}, \bibinfo{person}{Richard Benjamins}, \bibinfo{person}{Raja Chatila}, {and} \bibinfo{person}{Francisco Herrera}.} \bibinfo{year}{2020}\natexlab{}.
\newblock \showarticletitle{Explainable Artificial Intelligence ({XAI}): Concepts, taxonomies, opportunities and challenges toward responsible {AI}}.
\newblock \bibinfo{journal}{\emph{Information Fusion}}  \bibinfo{volume}{58} (\bibinfo{year}{2020}), \bibinfo{pages}{82--115}.
\newblock
\showISSN{1566-2535}
\newblock
\shownote{\url{https://doi.org/10.1016/j.inffus.2019.12.012}}.


\bibitem[Cabrera et~al\mbox{.}(2019)]%
        {cabrera2019fairvis}
\bibfield{author}{\bibinfo{person}{{\'A}ngel~Alexander Cabrera}, \bibinfo{person}{Will Epperson}, \bibinfo{person}{Fred Hohman}, \bibinfo{person}{Minsuk Kahng}, \bibinfo{person}{Jamie Morgenstern}, {and} \bibinfo{person}{Duen~Horng Chau}.} \bibinfo{year}{2019}\natexlab{}.
\newblock \showarticletitle{FairVis: Visual analytics for discovering intersectional bias in machine learning}. In \bibinfo{booktitle}{\emph{2019 IEEE Conference on Visual Analytics Science and Technology (VAST)}}. IEEE, \bibinfo{pages}{46--56}.
\newblock


\bibitem[Choi et~al\mbox{.}(2018)]%
        {choi2018stargan}
\bibfield{author}{\bibinfo{person}{Yunjey Choi}, \bibinfo{person}{Minje Choi}, \bibinfo{person}{Munyoung Kim}, \bibinfo{person}{Jung-Woo Ha}, \bibinfo{person}{Sunghun Kim}, {and} \bibinfo{person}{Jaegul Choo}.} \bibinfo{year}{2018}\natexlab{}.
\newblock \showarticletitle{StarGAN: Unified Generative Adversarial Networks for Multi-Domain Image-to-Image Translation}. In \bibinfo{booktitle}{\emph{Proceedings of the IEEE Conference on Computer Vision and Pattern Recognition}}.
\newblock


\bibitem[Colbois et~al\mbox{.}(2021)]%
        {colbois2021use}
\bibfield{author}{\bibinfo{person}{Laurent Colbois}, \bibinfo{person}{Tiago de Freitas~Pereira}, {and} \bibinfo{person}{S{\'e}bastien Marcel}.} \bibinfo{year}{2021}\natexlab{}.
\newblock \showarticletitle{On the use of automatically generated synthetic image datasets for benchmarking face recognition}. In \bibinfo{booktitle}{\emph{2021 IEEE International Joint Conference on Biometrics (IJCB)}}. IEEE, \bibinfo{pages}{1--8}.
\newblock


\bibitem[Denton et~al\mbox{.}(2019)]%
        {denton2019image}
\bibfield{author}{\bibinfo{person}{Emily Denton}, \bibinfo{person}{Ben Hutchinson}, \bibinfo{person}{Margaret Mitchell}, \bibinfo{person}{Timnit Gebru}, {and} \bibinfo{person}{Andrew Zaldivar}.} \bibinfo{year}{2019}\natexlab{}.
\newblock \showarticletitle{Image counterfactual sensitivity analysis for detecting unintended bias}.
\newblock \bibinfo{journal}{\emph{arXiv preprint arXiv:1906.06439}} (\bibinfo{year}{2019}).
\newblock


\bibitem[Diamond and Carey(1986)]%
        {diamond1986faces}
\bibfield{author}{\bibinfo{person}{Rhea Diamond} {and} \bibinfo{person}{Susan Carey}.} \bibinfo{year}{1986}\natexlab{}.
\newblock \showarticletitle{Why faces are and are not special: an effect of expertise.}
\newblock \bibinfo{journal}{\emph{Journal of experimental psychology: general}} \bibinfo{volume}{115}, \bibinfo{number}{2} (\bibinfo{year}{1986}), \bibinfo{pages}{107}.
\newblock


\bibitem[Engel et~al\mbox{.}(2018)]%
        {engel2018gansynth}
\bibfield{author}{\bibinfo{person}{Jesse Engel}, \bibinfo{person}{Kumar~Krishna Agrawal}, \bibinfo{person}{Shuo Chen}, \bibinfo{person}{Ishaan Gulrajani}, \bibinfo{person}{Chris Donahue}, {and} \bibinfo{person}{Adam Roberts}.} \bibinfo{year}{2018}\natexlab{}.
\newblock \showarticletitle{GANSynth: Adversarial Neural Audio Synthesis}. In \bibinfo{booktitle}{\emph{International Conference on Learning Representations}}.
\newblock


\bibitem[Goodfellow et~al\mbox{.}(2014a)]%
        {goodfellow2014generative}
\bibfield{author}{\bibinfo{person}{Ian Goodfellow}, \bibinfo{person}{Jean Pouget-Abadie}, \bibinfo{person}{Mehdi Mirza}, \bibinfo{person}{Bing Xu}, \bibinfo{person}{David Warde-Farley}, \bibinfo{person}{Sherjil Ozair}, \bibinfo{person}{Aaron Courville}, {and} \bibinfo{person}{Yoshua Bengio}.} \bibinfo{year}{2014}\natexlab{a}.
\newblock \showarticletitle{Generative adversarial nets}. In \bibinfo{booktitle}{\emph{Advances in neural information processing systems}}. \bibinfo{pages}{2672--2680}.
\newblock


\bibitem[Goodfellow et~al\mbox{.}(2014b)]%
        {goodfellow2014explaining}
\bibfield{author}{\bibinfo{person}{Ian~J Goodfellow}, \bibinfo{person}{Jonathon Shlens}, {and} \bibinfo{person}{Christian Szegedy}.} \bibinfo{year}{2014}\natexlab{b}.
\newblock \showarticletitle{Explaining and harnessing adversarial examples}.
\newblock \bibinfo{journal}{\emph{arXiv preprint arXiv:1412.6572}} (\bibinfo{year}{2014}).
\newblock


\bibitem[Guidotti et~al\mbox{.}(2018)]%
        {guidotti2018survey}
\bibfield{author}{\bibinfo{person}{Riccardo Guidotti}, \bibinfo{person}{Anna Monreale}, \bibinfo{person}{Franco Turini}, \bibinfo{person}{Dino Pedreschi}, {and} \bibinfo{person}{Fosca Giannotti}.} \bibinfo{year}{2018}\natexlab{}.
\newblock \bibinfo{title}{A Survey Of Methods For Explaining Black Box Models}.
\newblock
\newblock
\newblock
\shownote{\url{http://arxiv.org/abs/1802.01933}}.


\bibitem[H{\"a}rk{\"o}nen et~al\mbox{.}(2020)]%
        {harkonen2020ganspace}
\bibfield{author}{\bibinfo{person}{Erik H{\"a}rk{\"o}nen}, \bibinfo{person}{Aaron Hertzmann}, \bibinfo{person}{Jaakko Lehtinen}, {and} \bibinfo{person}{Sylvain Paris}.} \bibinfo{year}{2020}\natexlab{}.
\newblock \showarticletitle{GANspace: Discovering Interpretable GAN Controls}.
\newblock \bibinfo{journal}{\emph{Advances in neural information processing systems}}  \bibinfo{volume}{33} (\bibinfo{year}{2020}), \bibinfo{pages}{9841--9850}.
\newblock


\bibitem[Hendrycks et~al\mbox{.}(2021)]%
        {hendrycks2021natural}
\bibfield{author}{\bibinfo{person}{Dan Hendrycks}, \bibinfo{person}{Kevin Zhao}, \bibinfo{person}{Steven Basart}, \bibinfo{person}{Jacob Steinhardt}, {and} \bibinfo{person}{Dawn Song}.} \bibinfo{year}{2021}\natexlab{}.
\newblock \showarticletitle{Natural adversarial examples}. In \bibinfo{booktitle}{\emph{Proceedings of the IEEE/CVF Conference on Computer Vision and Pattern Recognition}}. \bibinfo{pages}{15262--15271}.
\newblock


\bibitem[Hohman et~al\mbox{.}(2018)]%
        {hohman2018visual}
\bibfield{author}{\bibinfo{person}{Fred Hohman}, \bibinfo{person}{Minsuk Kahng}, \bibinfo{person}{Robert Pienta}, {and} \bibinfo{person}{Duen~Horng Chau}.} \bibinfo{year}{2018}\natexlab{}.
\newblock \bibinfo{title}{Visual Analytics in Deep Learning: An Interrogative Survey for the Next Frontiers}.
\newblock
\newblock
\newblock
\shownote{\url{http://arxiv.org/abs/1801.06889}}.


\bibitem[Jahanian et~al\mbox{.}(2019)]%
        {jahanian2019steerability}
\bibfield{author}{\bibinfo{person}{Ali Jahanian}, \bibinfo{person}{Lucy Chai}, {and} \bibinfo{person}{Phillip Isola}.} \bibinfo{year}{2019}\natexlab{}.
\newblock \showarticletitle{On the''steerability" of generative adversarial networks}.
\newblock \bibinfo{journal}{\emph{arXiv preprint arXiv:1907.07171}} (\bibinfo{year}{2019}).
\newblock


\bibitem[Kang et~al\mbox{.}(2023)]%
        {kang2023gigagan}
\bibfield{author}{\bibinfo{person}{Minguk Kang}, \bibinfo{person}{Jun-Yan Zhu}, \bibinfo{person}{Richard Zhang}, \bibinfo{person}{Jaesik Park}, \bibinfo{person}{Eli Shechtman}, \bibinfo{person}{Sylvain Paris}, {and} \bibinfo{person}{Taesung Park}.} \bibinfo{year}{2023}\natexlab{}.
\newblock \showarticletitle{Scaling up GANs for Text-to-Image Synthesis}. In \bibinfo{booktitle}{\emph{Proceedings of the IEEE Conference on Computer Vision and Pattern Recognition (CVPR)}}.
\newblock


\bibitem[Karras et~al\mbox{.}(2019)]%
        {karras2019style}
\bibfield{author}{\bibinfo{person}{Tero Karras}, \bibinfo{person}{Samuli Laine}, {and} \bibinfo{person}{Timo Aila}.} \bibinfo{year}{2019}\natexlab{}.
\newblock \showarticletitle{A style-based generator architecture for generative adversarial networks}. In \bibinfo{booktitle}{\emph{Proceedings of the IEEE/CVF conference on computer vision and pattern recognition}}. \bibinfo{pages}{4401--4410}.
\newblock


\bibitem[Karras et~al\mbox{.}(2020)]%
        {Karras2020cvpr}
\bibfield{author}{\bibinfo{person}{Tero Karras}, \bibinfo{person}{Samuli Laine}, \bibinfo{person}{Miika Aittala}, \bibinfo{person}{Janne Hellsten}, \bibinfo{person}{Jaakko Lehtinen}, {and} \bibinfo{person}{Timo Aila}.} \bibinfo{year}{2020}\natexlab{}.
\newblock \showarticletitle{Analyzing and Improving the Image Quality of StyleGAN}. In \bibinfo{booktitle}{\emph{The IEEE/CVF Conference on Computer Vision and Pattern Recognition (CVPR)}}.
\newblock


\bibitem[Kingma and Dhariwal(2018)]%
        {kingma2018glow}
\bibfield{author}{\bibinfo{person}{Durk~P Kingma} {and} \bibinfo{person}{Prafulla Dhariwal}.} \bibinfo{year}{2018}\natexlab{}.
\newblock \showarticletitle{Glow: Generative flow with invertible 1x1 convolutions}.
\newblock \bibinfo{journal}{\emph{Advances in neural information processing systems}}  \bibinfo{volume}{31} (\bibinfo{year}{2018}).
\newblock


\bibitem[Kingma and Welling(2013)]%
        {kingma2013auto}
\bibfield{author}{\bibinfo{person}{Diederik~P Kingma} {and} \bibinfo{person}{Max Welling}.} \bibinfo{year}{2013}\natexlab{}.
\newblock \showarticletitle{Auto-encoding variational bayes}.
\newblock \bibinfo{journal}{\emph{arXiv preprint arXiv:1312.6114}} (\bibinfo{year}{2013}).
\newblock


\bibitem[Lipton(2016)]%
        {lipton2016mythos}
\bibfield{author}{\bibinfo{person}{Zachary~Chase Lipton}.} \bibinfo{year}{2016}\natexlab{}.
\newblock \bibinfo{title}{The Mythos of Model Interpretability}.
\newblock
\newblock
\newblock
\shownote{\url{http://arxiv.org/abs/1606.03490}}.


\bibitem[Liu et~al\mbox{.}(2018)]%
        {liu2018analyzing}
\bibfield{author}{\bibinfo{person}{Mengchen Liu}, \bibinfo{person}{Shixia Liu}, \bibinfo{person}{Hang Su}, \bibinfo{person}{Kelei Cao}, {and} \bibinfo{person}{Jun Zhu}.} \bibinfo{year}{2018}\natexlab{}.
\newblock \showarticletitle{Analyzing the Noise Robustness of Deep Neural Networks}. In \bibinfo{booktitle}{\emph{2018 IEEE Conference on Visual Analytics Science and Technology (VAST)}}. \bibinfo{publisher}{IEEE}, \bibinfo{address}{New York City, USA}, \bibinfo{pages}{60--71}.
\newblock
\urldef\tempurl%
\url{https://doi.org/10.1109/VAST.2018.8802509}
\showDOI{\tempurl}


\bibitem[Ma et~al\mbox{.}(2019)]%
        {ma2019explaining}
\bibfield{author}{\bibinfo{person}{Yuxin Ma}, \bibinfo{person}{Tiankai Xie}, \bibinfo{person}{Jundong Li}, {and} \bibinfo{person}{Ross Maciejewski}.} \bibinfo{year}{2019}\natexlab{}.
\newblock \showarticletitle{Explaining vulnerabilities to adversarial machine learning through visual analytics}.
\newblock \bibinfo{journal}{\emph{IEEE transactions on visualization and computer graphics}} \bibinfo{volume}{26}, \bibinfo{number}{1} (\bibinfo{year}{2019}), \bibinfo{pages}{1075--1085}.
\newblock


\bibitem[Ning et~al\mbox{.}(2019)]%
        {ning2019review}
\bibfield{author}{\bibinfo{person}{Yishuang Ning}, \bibinfo{person}{Sheng He}, \bibinfo{person}{Zhiyong Wu}, \bibinfo{person}{Chunxiao Xing}, {and} \bibinfo{person}{Liang-Jie Zhang}.} \bibinfo{year}{2019}\natexlab{}.
\newblock \showarticletitle{A review of deep learning based speech synthesis}.
\newblock \bibinfo{journal}{\emph{Applied Sciences}} \bibinfo{volume}{9}, \bibinfo{number}{19} (\bibinfo{year}{2019}), \bibinfo{pages}{4050}.
\newblock


\bibitem[Pidhorskyi et~al\mbox{.}(2020)]%
        {pidhorskyi2020adversarial}
\bibfield{author}{\bibinfo{person}{Stanislav Pidhorskyi}, \bibinfo{person}{Donald~A Adjeroh}, {and} \bibinfo{person}{Gianfranco Doretto}.} \bibinfo{year}{2020}\natexlab{}.
\newblock \showarticletitle{Adversarial Latent Autoencoders}. In \bibinfo{booktitle}{\emph{Proceedings of the IEEE Computer Society Conference on Computer Vision and Pattern Recognition (CVPR)}}. \bibinfo{pages}{14104--14113}.
\newblock


\bibitem[Powers(2020)]%
        {powers2020evaluation}
\bibfield{author}{\bibinfo{person}{David~MW Powers}.} \bibinfo{year}{2020}\natexlab{}.
\newblock \showarticletitle{Evaluation: from precision, recall and F-measure to ROC, informedness, markedness and correlation}.
\newblock \bibinfo{journal}{\emph{arXiv preprint arXiv:2010.16061}} (\bibinfo{year}{2020}).
\newblock


\bibitem[Ramesh et~al\mbox{.}(2021)]%
        {ramesh2021zero}
\bibfield{author}{\bibinfo{person}{Aditya Ramesh}, \bibinfo{person}{Mikhail Pavlov}, \bibinfo{person}{Gabriel Goh}, \bibinfo{person}{Scott Gray}, \bibinfo{person}{Chelsea Voss}, \bibinfo{person}{Alec Radford}, \bibinfo{person}{Mark Chen}, {and} \bibinfo{person}{Ilya Sutskever}.} \bibinfo{year}{2021}\natexlab{}.
\newblock \showarticletitle{Zero-shot text-to-image generation}. In \bibinfo{booktitle}{\emph{International Conference on Machine Learning}}. PMLR, \bibinfo{pages}{8821--8831}.
\newblock


\bibitem[Ren et~al\mbox{.}(2020)]%
        {ren2020fastspeech}
\bibfield{author}{\bibinfo{person}{Yi Ren}, \bibinfo{person}{Chenxu Hu}, \bibinfo{person}{Xu Tan}, \bibinfo{person}{Tao Qin}, \bibinfo{person}{Sheng Zhao}, \bibinfo{person}{Zhou Zhao}, {and} \bibinfo{person}{Tie-Yan Liu}.} \bibinfo{year}{2020}\natexlab{}.
\newblock \showarticletitle{Fastspeech 2: Fast and high-quality end-to-end text to speech}.
\newblock \bibinfo{journal}{\emph{arXiv preprint arXiv:2006.04558}} (\bibinfo{year}{2020}).
\newblock


\bibitem[Rezende and Mohamed(2015)]%
        {rezende2015variational}
\bibfield{author}{\bibinfo{person}{Danilo Rezende} {and} \bibinfo{person}{Shakir Mohamed}.} \bibinfo{year}{2015}\natexlab{}.
\newblock \showarticletitle{Variational inference with normalizing flows}. In \bibinfo{booktitle}{\emph{International conference on machine learning}}. PMLR, \bibinfo{pages}{1530--1538}.
\newblock


\bibitem[Rombach et~al\mbox{.}(2022)]%
        {rombach2022high}
\bibfield{author}{\bibinfo{person}{Robin Rombach}, \bibinfo{person}{Andreas Blattmann}, \bibinfo{person}{Dominik Lorenz}, \bibinfo{person}{Patrick Esser}, {and} \bibinfo{person}{Bj{\"o}rn Ommer}.} \bibinfo{year}{2022}\natexlab{}.
\newblock \showarticletitle{High-resolution image synthesis with latent diffusion models}. In \bibinfo{booktitle}{\emph{Proceedings of the IEEE/CVF conference on computer vision and pattern recognition}}. \bibinfo{pages}{10684--10695}.
\newblock


\bibitem[Schroff et~al\mbox{.}(2015)]%
        {schroff2015facenet}
\bibfield{author}{\bibinfo{person}{Florian Schroff}, \bibinfo{person}{Dmitry Kalenichenko}, {and} \bibinfo{person}{James Philbin}.} \bibinfo{year}{2015}\natexlab{}.
\newblock \showarticletitle{Facenet: A unified embedding for face recognition and clustering}. In \bibinfo{booktitle}{\emph{Proceedings of the IEEE conference on computer vision and pattern recognition}}. \bibinfo{pages}{815--823}.
\newblock


\bibitem[Serengil and Ozpinar(2020)]%
        {serengil2020lightface}
\bibfield{author}{\bibinfo{person}{Sefik~Ilkin Serengil} {and} \bibinfo{person}{Alper Ozpinar}.} \bibinfo{year}{2020}\natexlab{}.
\newblock \showarticletitle{LightFace: A Hybrid Deep Face Recognition Framework}. In \bibinfo{booktitle}{\emph{2020 Innovations in Intelligent Systems and Applications Conference (ASYU)}}. IEEE, \bibinfo{pages}{23--27}.
\newblock
\urldef\tempurl%
\url{https://doi.org/10.1109/ASYU50717.2020.9259802}
\showDOI{\tempurl}


\bibitem[Shen et~al\mbox{.}(2020)]%
        {shen2020interpreting}
\bibfield{author}{\bibinfo{person}{Yujun Shen}, \bibinfo{person}{Jinjin Gu}, \bibinfo{person}{Xiaoou Tang}, {and} \bibinfo{person}{Bolei Zhou}.} \bibinfo{year}{2020}\natexlab{}.
\newblock \showarticletitle{Interpreting the Latent Space of GANs for Semantic Face Editing}. In \bibinfo{booktitle}{\emph{CVPR}}.
\newblock


\bibitem[Shimizu et~al\mbox{.}(2022)]%
        {shimizu2022human}
\bibfield{author}{\bibinfo{person}{Kye Shimizu}, \bibinfo{person}{Naoto Ienaga}, \bibinfo{person}{Kazuma Takada}, \bibinfo{person}{Maki Sugimoto}, {and} \bibinfo{person}{Shunichi Kasahara}.} \bibinfo{year}{2022}\natexlab{}.
\newblock \showarticletitle{Human Latent Metrics: Perceptual and Cognitive Response Correlates to Distance in GAN Latent Space for Facial Images}. In \bibinfo{booktitle}{\emph{ACM Symposium on Applied Perception 2022}}. \bibinfo{pages}{1--10}.
\newblock


\bibitem[Terh{\"o}rst et~al\mbox{.}(2021)]%
        {terhorst2021comprehensive}
\bibfield{author}{\bibinfo{person}{Philipp Terh{\"o}rst}, \bibinfo{person}{Jan~Niklas Kolf}, \bibinfo{person}{Marco Huber}, \bibinfo{person}{Florian Kirchbuchner}, \bibinfo{person}{Naser Damer}, \bibinfo{person}{Aythami~Morales Moreno}, \bibinfo{person}{Julian Fierrez}, {and} \bibinfo{person}{Arjan Kuijper}.} \bibinfo{year}{2021}\natexlab{}.
\newblock \showarticletitle{A comprehensive study on face recognition biases beyond demographics}.
\newblock \bibinfo{journal}{\emph{IEEE Transactions on Technology and Society}} \bibinfo{volume}{3}, \bibinfo{number}{1} (\bibinfo{year}{2021}), \bibinfo{pages}{16--30}.
\newblock


\bibitem[Topol(2019)]%
        {topol2019high}
\bibfield{author}{\bibinfo{person}{Eric~J Topol}.} \bibinfo{year}{2019}\natexlab{}.
\newblock \showarticletitle{High-performance medicine: the convergence of human and artificial intelligence}.
\newblock \bibinfo{journal}{\emph{Nature medicine}} \bibinfo{volume}{25}, \bibinfo{number}{1} (\bibinfo{year}{2019}), \bibinfo{pages}{44--56}.
\newblock


\bibitem[Vahdat and Kautz(2020)]%
        {vahdat2020nvae}
\bibfield{author}{\bibinfo{person}{Arash Vahdat} {and} \bibinfo{person}{Jan Kautz}.} \bibinfo{year}{2020}\natexlab{}.
\newblock \showarticletitle{NVAE: A deep hierarchical variational autoencoder}.
\newblock \bibinfo{journal}{\emph{Advances in neural information processing systems}}  \bibinfo{volume}{33} (\bibinfo{year}{2020}), \bibinfo{pages}{19667--19679}.
\newblock


\bibitem[Wang et~al\mbox{.}(2021)]%
        {wang2021attribute}
\bibfield{author}{\bibinfo{person}{Rui Wang}, \bibinfo{person}{Jian Chen}, \bibinfo{person}{Gang Yu}, \bibinfo{person}{Li Sun}, \bibinfo{person}{Changqian Yu}, \bibinfo{person}{Changxin Gao}, {and} \bibinfo{person}{Nong Sang}.} \bibinfo{year}{2021}\natexlab{}.
\newblock \showarticletitle{Attribute-specific control units in stylegan for fine-grained image manipulation}. In \bibinfo{booktitle}{\emph{Proceedings of the 29th ACM International Conference on Multimedia}}. \bibinfo{pages}{926--934}.
\newblock


\bibitem[Yager and Dunstone(2008)]%
        {yager2008biometric}
\bibfield{author}{\bibinfo{person}{Neil Yager} {and} \bibinfo{person}{Ted Dunstone}.} \bibinfo{year}{2008}\natexlab{}.
\newblock \showarticletitle{The biometric menagerie}.
\newblock \bibinfo{journal}{\emph{IEEE transactions on pattern analysis and machine intelligence}} \bibinfo{volume}{32}, \bibinfo{number}{2} (\bibinfo{year}{2008}), \bibinfo{pages}{220--230}.
\newblock


\bibitem[Yapo and Weiss(2018)]%
        {yapo2018ethical}
\bibfield{author}{\bibinfo{person}{Adrienne Yapo} {and} \bibinfo{person}{Joseph Weiss}.} \bibinfo{year}{2018}\natexlab{}.
\newblock \showarticletitle{Ethical implications of bias in machine learning}. In \bibinfo{booktitle}{\emph{Proceedings of the 51st Hawaii International Conference on System Sciences}}.
\newblock


\bibitem[Zhang et~al\mbox{.}(2023)]%
        {zhang2023generating}
\bibfield{author}{\bibinfo{person}{Jianrong Zhang}, \bibinfo{person}{Yangsong Zhang}, \bibinfo{person}{Xiaodong Cun}, \bibinfo{person}{Shaoli Huang}, \bibinfo{person}{Yong Zhang}, \bibinfo{person}{Hongwei Zhao}, \bibinfo{person}{Hongtao Lu}, {and} \bibinfo{person}{Xi Shen}.} \bibinfo{year}{2023}\natexlab{}.
\newblock \showarticletitle{T2M-GPT: Generating Human Motion from Textual Descriptions with Discrete Representations}. In \bibinfo{booktitle}{\emph{Proceedings of the IEEE/CVF Conference on Computer Vision and Pattern Recognition (CVPR)}}.
\newblock


\bibitem[Zhang et~al\mbox{.}(2018)]%
        {zhang2018perceptual}
\bibfield{author}{\bibinfo{person}{Richard Zhang}, \bibinfo{person}{Phillip Isola}, \bibinfo{person}{Alexei~A Efros}, \bibinfo{person}{Eli Shechtman}, {and} \bibinfo{person}{Oliver Wang}.} \bibinfo{year}{2018}\natexlab{}.
\newblock \showarticletitle{The Unreasonable Effectiveness of Deep Features as a Perceptual Metric}. In \bibinfo{booktitle}{\emph{CVPR}}.
\newblock


\end{thebibliography}


\end{document}